\documentclass[12pt,preprint,flushrt]{aastex}
\begin{document}
\newcommand{\etal}{{\it et al.}}
\include{idl}

\title{Phase Coherent Timing of RX J0806.3+1527 with ROSAT and CHANDRA}
\author{Tod E. Strohmayer}
\affil{Laboratory for High Energy Astrophysics, NASA's Goddard Space Flight 
Center, Greenbelt, MD 20771; stroh@clarence.gsfc.nasa.gov}

\begin{abstract}

RX J0806.3+1527 is an ultra-compact, double degenerate binary with the
shortest known orbital period (321.5 s).  Hakala et al. (2003) have
recently reported new optical measurements of the orbital frequency of
the source which indicate that the frequency has increased over the
$\approx 9$ years since the earliest ROSAT observations.  They find
two candidate solutions for the long term change in the frequency;
$\dot\nu \approx 3$ or $6\times 10^{-16}$ Hz s$^{-1}$. Here we present
the results of a phase coherent timing study of the archival ROSAT and
Chandra data for RX J0806.3+1527 in the light of these new
constraints.  We find that the ROSAT -- Chandra timing data are
consistent with both of the solutions reported by Hakala et al., but
that the higher $\dot\nu = 6.1 \times 10^{-16}$ Hz s$^{-1}$ solution
is favored at the $\approx 97\%$ level. This large a $\dot\nu$ can be
accomodated by an $\approx 1 M_{\odot}$ detached double degenerate
system powered in the X-ray by electrical energy (Wu et al. 2002).
With such a large $\dot\nu$ the system provides a unique opportunity
to explore the interaction of gravitational radiation and
electromagnetic torques on the evolution of an ultracompact binary.

\end{abstract}


\section{Introduction}

Double degenerate systems containing a pair of interacting white
dwarfs are the most compact binaries known. Evidence has been mounting
recently that the two systems with the shortest inferred orbital
periods, RX J1914.4+2456 (569 s, hereafter J1914) and RX J0806.3+1527
(321 s, hereafter J0806), may represent a new class of double
degenerates whose X-ray flux is powered by a unipolar induction
process rather than accretion (see Wu et al. 2002; Strohmayer 2002;
and Hakala et al. 2003). These objects have much in common (see
Cropper et al. 2003 for a recent review). Their X-ray lightcurves show
100\% modulation with a sharp rise and more gradual decay, suggestive
of a small X-ray emitting region on the primary (Cropper et
al. 1998). In both systems the optical maximum lags the X-ray maximum
by about 0.6 of a cycle, consistent with the optical variations
resulting from X-ray heating of the secondary (Ramsay et al.  2000;
Israel et al. 2003).

Of the models that have been proposed for J1914 and J0806, the
Intermediate Polar (IP) scenario is looking increasingly unlikely (see
Cropper et al. 2003 for a discussion). All the remaining models are
variants of a double degenerate scenario. In the Polar model the
primary is magnetic, the binary is synchronized and the X-ray flux is
accretion driven (Cropper et al. 1998). Some difficulties with this
model are the lack of circular polarization in the optical (although
magnetic fields weak enough to preclude detection of polarization may
still be able to enforce synchronization), and the lack of any hard
($> 2$ keV) spectral component, as is usually seen in Polars. Another
variant is the direct accretor model (Marsh \& Steeghs 2002; Ramsay et
al. 2002). In this model both components are non-magnetic, and the
accretion stream impacts directly onto the surface of the primary
(i.e. without the formation of an accretion disk). Shortcomings of
this model include the stable X-ray to optical phasing of the
lightcurves, and the need for a relatively low mass ($< 0.5
M_{\odot}$) primary.

An important constraint on the models can be obtained by measuring the
orbital evolution in these systems.  If the donors are degenerate,
then stable accretion will lead to a widening of the orbit and a
decrease in the orbital frequency, $\dot\nu$. From a timing study of
archival ROSAT and ASCA data Strohmayer (2002) found evidence that the
orbital frequency of J1914 is increasing at a rate consistent with
loss of orbital angular momentum to gravitational radiation. Recently,
Hakala et al. (2003) have used archival ROSAT and new optical timing
measurements of J0806 to constrain the orbital evolution. They find
that, similarly to J1914, the orbital frequency of J0806 is also
increasing, but at a rate that is factor of about 30 or more higher
than for J1914.  Considering the more compact orbit in J0806, and the
strong $\nu^{11/3}$ orbital frequency dependence of gravitational
radiation, a substantially higher $\dot\nu$ is not unexpected.

Hakala et al. (2003) used period measurements from ROSAT in 1994 -
1995 as well as more recent optical data from the ESO VLT and the
Nordic Optical Telescope (NOT) on La Palma to explore the orbital
evolution of J0806. They derived a frequency history at three epochs
and deduced a frequency increase.  Here we report the results of a
coherent, phase timing study of J0806 using the 1994 - 1995 ROSAT data
combined with the public Chandra observations from November,
2001. Hakala et al. (2003) did not use the Chandra data in their study
because it is too short in duration for a precision frequency
measurement. By themselves the ROSAT and Chandra data are insufficient
to unambiguously constrain the frequency evolution because of their
sparseness. However, in the light of the additional constraints on
$\dot\nu$ provided by the Hakala et al. study, we can use phase timing
analysis of the ROSAT and Chandra data to test for consistency with
and to see if the data favor one of the possible solutions found by
Hakala et al. (2003).  We find that the ROSAT and Chandra phase timing
data are consistent with the two possible $\dot\nu$ solutions found by
Hakala et al. (2003), and that the solution with the larger $\dot\nu
\approx 6 \times 10^{-16}$ Hz s$^{-1}$ provides a better fit to the
phase timing data and is preferred at about the $97\%$ confidence
level. We also briefly discuss the implications of this finding for
the models for J0806 and its cousin source J1914.

\section{Data Extraction and Analysis}

The ROSAT HRI observed J0806 for a total of $\approx$ 16 ksec in the
interval from October, 1994 through April, 1995. Burwitz \& Reinsch
(2001) reported on the pulse timing from these observations.  Chandra
observed J0806 for 20 ksec in November, 2001. Spectroscopy from this
observation is reported by Israel et al. (2003).  All these data are
now in the public domain.  We used standard HEASARC FTOOLS data
analysis packages (i.e., XSELECT) to extract and analyse the data.  We
began by producing images and extracting source events. We next
applied barycentric corrections to the event times for each
observation. We used the standard mission FTOOLS in conjunction with
the ROSAT and Chandra orbital and JPL (DE200) solar system ephemerides
(Standish et al. 1992). For the source position we used the
coordinates, ($\alpha = 08^h 06^m 23.2^s$, $\delta = 15^{\circ} 27'
30.2''$ :J2000), obtained by Israel et al. (2002) from their study of
the optical-infrared counterpart to J0806. Our analysis resulted in a
total of 725 and 6076 photons from the ROSAT and Chandra observations,
respectively.

\subsection{Coherent Timing Methods}

We performed our coherent timing studies using the $Z_n^2$ statistic
(Buccheri 1983; see also Strohmayer 2002 and Strohmayer \& Markwardt
2002 for examples of the use of this statistic in a similar
context). We use a two parameter phase model, with $\phi (t) = \nu_0
(t-t_0) + \frac{1}{2} \dot\nu (t - t_0)^2$. Since the methods are
described elsewhere we do not repeat the details here. From their
study of the ROSAT data, Burwitz \& Reinsch (2001) found two candidate
periods, 321.5393 or 321.5465 s, with an uncertainty of 0.4 ms. We
began by first analyzing just the ROSAT events and confirm these
results. In terms of frequencies they are 3.1100392 and 3.10997285
mHz, respectively, with an uncertainty of about 4 nHz.

From their analysis of the orbital frequency at three different epochs
spanning about 9.2 years Hakala et al. (2003) found two possible
solutions for $\dot\nu$.  Two solutions were possible because of the
ambiguity in the period obtained from the ROSAT observations.  We used
the ROSAT and Chandra data to perform a grid search in the $\nu -
\dot\nu$ plane in the vicinity of these two possible solutions in
order to determine whether the combined ROSAT -- Chandra data are
consistent with them and to see if perhaps one of them is favored.  We
calculated a $\chi^2$ statistic by comparing the model predicted
phases with the observed phases (see Strohmayer 2002 for details). We
had a total of 31 phase measurements, 10 from the ROSAT epoch and 21
from the Chandra data, giving us fits with 29 degrees of freedom. The
results of our grid search are summarized in Figure 1. Panel (a) shows
the $1\sigma$ confidence contours in the vicinity of $\dot\nu = 3.1
\times 10^{-16}$ Hz s$^{-1}$, while panel (b) shows the same
quantities in the vicinity of the larger $\dot\nu$ solution, $\dot\nu
= 6.1 \times 10^{-16}$ Hz s$^{-1}$. In each figure the best solution
and errors from the Hakala et al. study are marked by the solid and
dashed horizontal lines, respectively. We find that the ROSAT --
Chandra timing data have solutions which are consistent with both of
the Hakala et al. (2003) solutions. However, we also find that the
$\dot\nu = 6.11 \times 10^{-16}$ Hz s$^{-1}$ gives a better fit. This
solution has a minimum $\chi^2 = 33.1$, while the other $\dot\nu$
solution has a minimum $\chi^2 = 40.8$. The $\Delta\chi^2 = 7.7$
between the two solutions favors the higher $\dot\nu$ solution at
about the $97\%$ level. Although this level of significance does not
provide concrete proof, it is suggestive that the higher $\dot\nu$
solution is the correct one.

Figures 2a and 2b compare the phase residuals computed from the models
with $\dot\nu = 3.1 \times 10^{-16}$ and $6.1 \times 10^{-16}$ Hz
s$^{-1}$, respectively.  Interestingly, much of the improvement in the
fit results from the improvement of the residuals to the ROSAT
data. Although this may at first seem surprising, these data span an
interval of about 7 months. A $\dot\nu$ of $6 \times 10^{-16}$ Hz
s$^{-1}$ produces a phase advance on the order of $\approx 0.2$ cycles
over such a time span, so that the ROSAT data are in principle
sensitive to a $\dot\nu$ of this size. We note that without the {\it a
priori} constraints on $\dot\nu$ provided by the Hakala et
al. results, the ROSAT -- Chandra data can be fit by a number of
different combinations of $\nu$ and $\dot\nu$. Some of these ``alias''
solutions can be seen in Figure 1a and b. 

\subsection{Discussion and Implications}

The observations by Hakala et al. (2003) of the orbital frequency
history of J0806 provide strong evidence that the orbit is decaying at
a rate as predicted by the ultra-compact binary model in which the
loss of orbital angular momentum is dominated by gravitational
radiation and there is no mass exchange between the two stars.  We
have shown that phase coherent timing of the ROSAT and Chandra data
are consistent with the rate of orbital frequency increase deduced by
Hakala et al (2003), and that the ROSAT - Chandra data provide
suggestive evidence that the higher $\dot\nu$ value is the correct
one.

For a detached binary with a circular orbit the rate of change of the
orbital frequency due to gravitational radiation is (see for example,
Evans, Iben \& Smarr 1987; Taylor \& Weisberg 1989)
\begin{equation} 
\dot\nu_{gr} = 1.64 \times 10^{-17} \ \left ( \frac{\nu}{10^{-3}\; 
{\rm Hz}} \right )^5 \; \left ( \frac{\mu}{M_{\odot}} \right ) \; 
\left ( \frac{a}{10^{10} \; {\rm cm}} \right )^2 \;\; {\rm Hz} 
\ {\rm s}^{-1} \; ,
\end{equation}
where, $\nu$, $\mu$, and $a$ are the orbital frequency, reduced mass
and orbital separation of the components, respectively. If the orbital
decay results only from gravitational radiation losses and there is no
mass transfer, then the constraint on $\dot\nu$ implies a constraint
on the so called ``chirp mass,''
\begin{equation} 
\left ( \frac{M_{ch}}{M_{\odot}} \right )^{5/3}  = 
\left ( \frac{\mu}{M_{\odot}} \right ) \ \left ( 
\frac{m_1 + m_2}{M_{\odot}} \right )^{2/3}  = 2.7 \times 10^{16} \left ( 
\frac{\nu}{10^{-3} \ {\rm Hz}} \right )^{-11/3} \ \dot\nu \; .
\end{equation}
This constraint follows directly from equation (1) and the use of
Kepler's law to substitute for the orbital separation, $a$.  This
expression also assumes no spin - orbit coupling. If the stars remain
synchronized with the orbit, then the expression is modified to
account for the angular momentum required to spin up the components as
the orbit shrinks.  We can use these expressions to constrain the
component masses of J0806 under the assumption that the orbital
dissipation is dominated by gravitational radiation. Figure 3 shows
the $1\sigma$ contours in the mass plane for the two possible
$\dot\nu$ solutions. Contours calculated assuming no spin - orbit
coupling (solid curves), and full synchronization (dashed curves) are
also shown.  The dotted lines mark the region where the mass ratio, $q
= m_{sec} / m_{prim}$ is between 0.5 and 1. Binary evolution
calculations for ultra-compact, double degenerate systems suggest that
they preferrentially form with mass ratios $\sim 1$ (see Nelemans et
al. 2001, Napiwotzki et al. 2002). The inferred masses are consistent
with double degenerate system with a total system mass $\sim 1
M_{\odot}$ (see also Hakala et al. 2003).

The inferred rate of orbital evolution in J0806 is so large that it
opens up a unique opportunity for phase timing observations to track
its evolution in detail.  For example, at a rate of $6 \times
10^{-16}$ Hz s$^{-1}$ the phase will advance by $\approx 0.3$ cycles
in only a year. A glance at Figure 2 confirms that such changes can
be easily measured with Chandra in less than a year.  

The inferred spin-up of the system is difficult to understand in the
context of an accreting double-degenerate system.  Stable accretion
requires that the orbital separation increases, contrary to the
observations. (see for example, Rappaport, Joss \& Webbink 1982;
Nelemans et al. 2001).  As noted by Strohmayer (2002) and Hakala et
al. (2003), the observation of spin-up in J1914 and J0806 can be
accomodated in the unipolar induction, or electric star model of Wu et
al. (2002). Indeed, for the large $\dot\nu$ inferred from J0806, it is
possible that a majority of the orbital dissipation may arise from the
electromagnetic torques.  It is likely that instabilities in this
process as, for example, caused by induced magnetic fields, will
introduce variability both in the X-ray flux and the electromagnetic
torque (Cropper et al. 2003). With precise, phase coherent timing, it
is possible that such variations could be detected, and the electric
star model explored in detail.  As noted by Hakala et al. (2003), this
would also be possible with independent optically derived constraints
on the component masses, and would also be facilitated by future
gravitational wave measurements (with, for example LISA).

\centerline{\bf References}

\noindent{} Buccheri, R. et al. 1983, A\&A, 128, 245

\noindent{} Burwitz, V. \& Reinsch, K. 2001, X-ray Astronomy: stellar
endpoints, AGN, and the diffuse X-ray background, Bologna, Italy, eds.
White, N. E., Malaguti, G., and Palumbo, G., AIP conference
proceedings, 599, 522.

\noindent{} Cropper, M., Ramsay, G., Wu, K. \& Hakala, P. 2003, in
Proc. Third Workshop on Magnetic CVs, Cape Town, (astro-ph/0302240).

\noindent{} Cropper, M. et al. 1998, MNRAS, 293, L57.

\noindent{} Evans, C. R., Iben, I. \& Smarr, L. 1987, ApJ, 323, 129.

\noindent{} Hakala, P. et al. 2003, MNRAS, in press (astro-ph/0305283).

\noindent{} Iben, I. \& Tutukov, A. V. 1991, ApJ, 370, 615.

\noindent{} Israel, G. L. et al. 2003, in Proc. Third
Workshop on Magnetic CVs, Cape Town, (astro-ph/0303124).

\noindent{} Israel, G. L. et al. 2002, A\&A, 386, L131.

\noindent{} Israel, G. L. et al. 1999, A\&A, 349, L1.

\noindent{} Marsh, T. R. \& Steeghs, D. 2002, MNRAS, 331, L7.

\noindent{} Motch, C. et al. 1996, A\&A, 307, 459.

\noindent{} Napiwotzki, R. et al. 2002, A\&A, 386, 957.

\noindent{} Nelemans, G., Portegies Zwart, S. F., Verbunt, F. \& Yungelson, 
L. R. 2001a, A\&A, 368, 939.

\noindent{} Ramsay, G., Cropper, M., Wu, K., Mason, K.~O., \& Hakala, P.\ 
2000, MNRAS, 311, 75.

\noindent{} Ramsay, G., et al. 2002, MNRAS, 333, 575.

% \noindent{} Ramsay, G. Hakala, P. \& Cropper, M. 2002, MNRAS, 332, L7.

\noindent{} Rappaport, S., Joss, P. C. \& Webbink, R. F. 1982, ApJ, 254, 616.

\noindent{} Standish, E. M., Newhall, X. X., Williams, J. G. \& Yeomans, D. K
            1992, in Explanatory Supplememt to the Astronomical Almanac, ed. 
            P. K. Seidelmann (Mill Valley: University Science), 239.

\noindent{} Strohmayer, T. E. 2002, ApJ, 581, 577.

\noindent{} Strohmayer, T. E., \& Markwardt, C. B., 2002, ApJ, 577, 337.

\noindent{} Strohmayer, T. E., \& Markwardt, C. B., 1999, ApJ, 516, L81.

\noindent{} Taylor, J. H. \& Weisberg, J. M. 1989, ApJ, 345, 434.

\noindent{} Wu, K., Cropper, M., Ramsay, G. \& Sekiguchi, K. 2002, MNRAS, 
331, 221.

\pagebreak
\centerline{\bf Figure Captions}

\figcaption[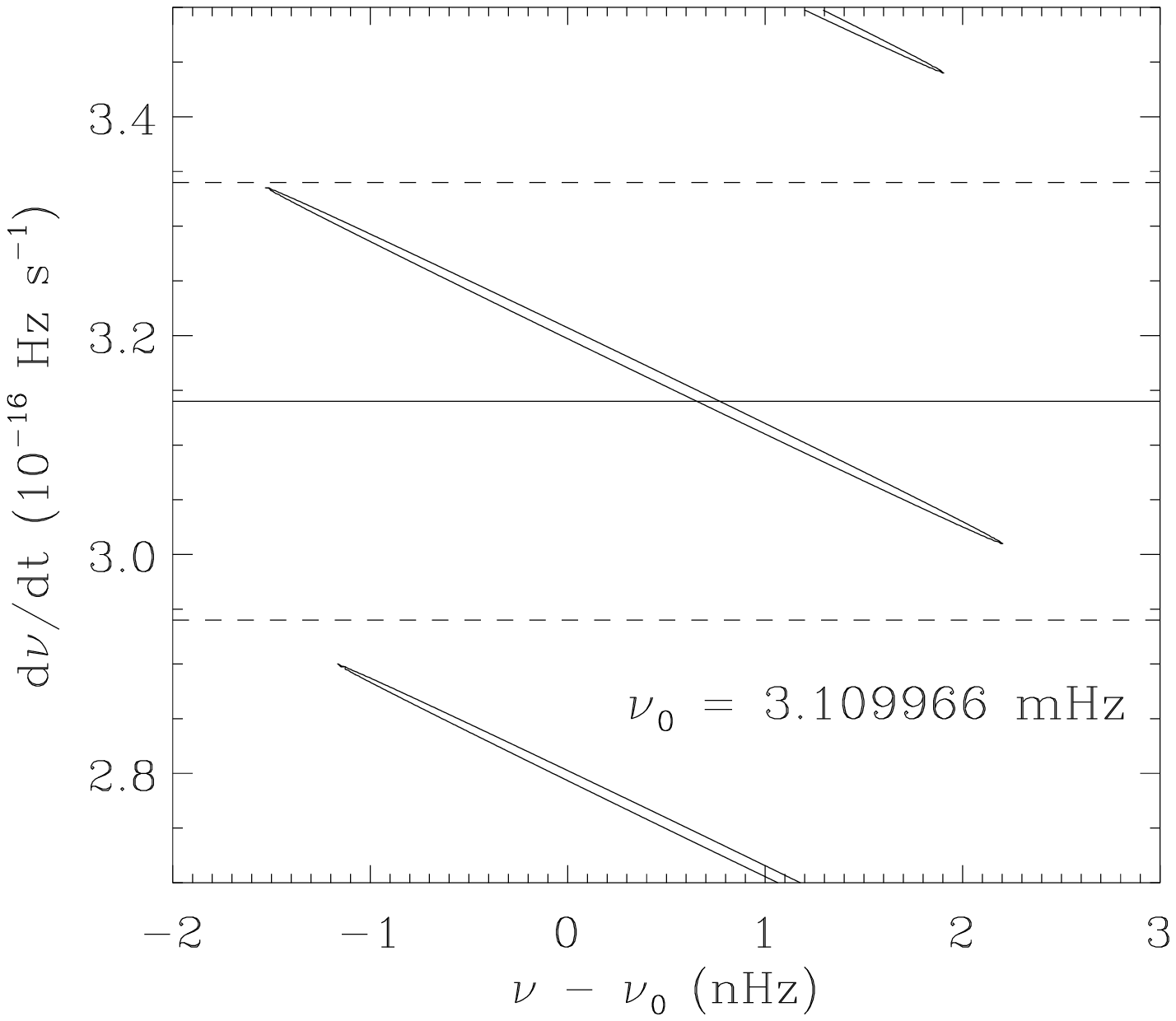, 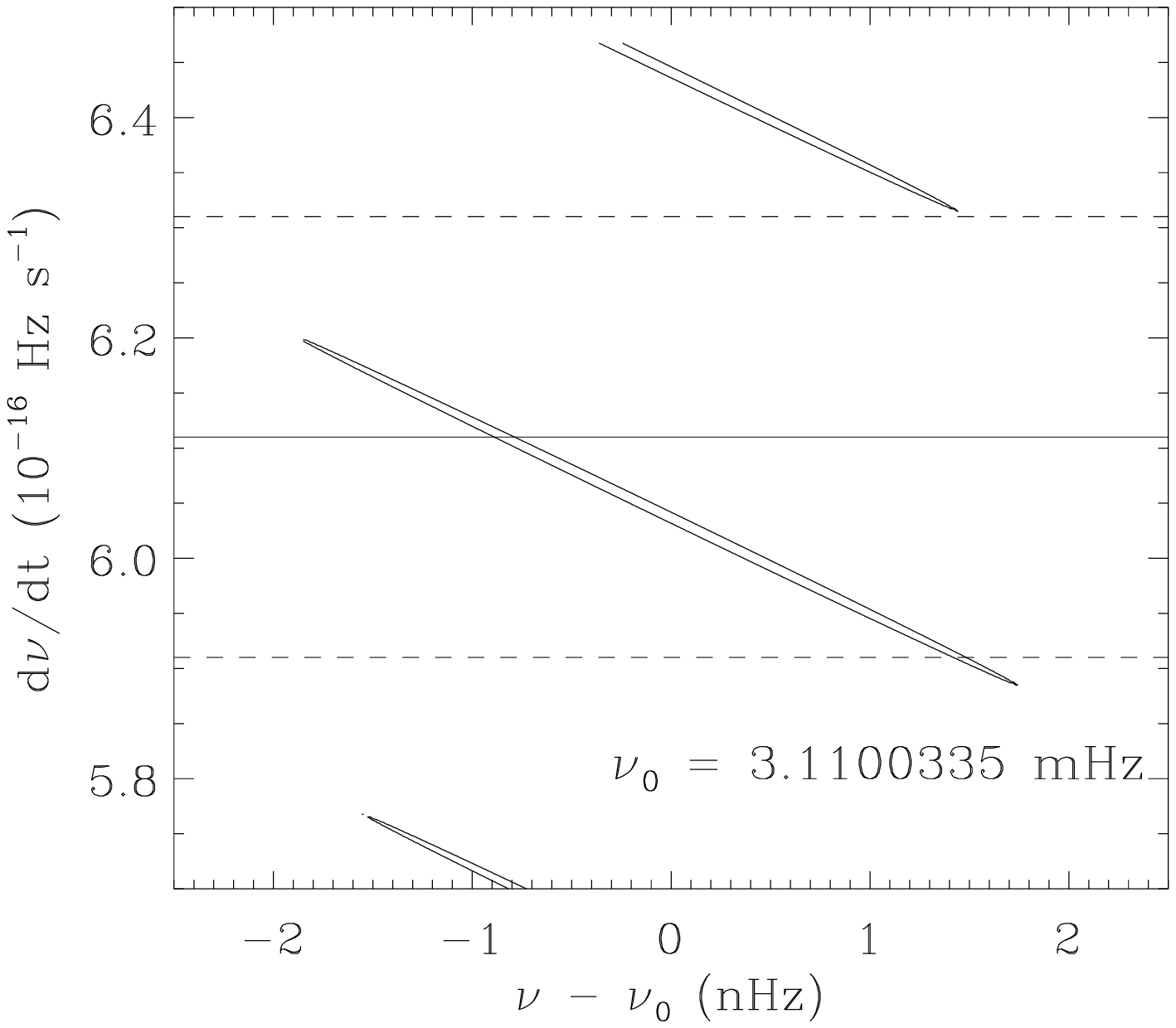]{Results of our ROSAT -- Chandra timing
study for J0806. (a) The $1\sigma$ confidence contours in the $\nu$,
$\dot\nu$ plane are shown in the vicinity of the lower $\dot\nu
\approx 3\times 10^{-16}$ Hz s$^{-1}$ solution of Hakala et
al. (2003). The best $\chi^2$ for this solution was $\chi^2_{min} =
40.8$. The Hakala et al. solution and $1\sigma$ errors are marked by
the horizontal solid and dashed lines, respectively. The nearby,
``alias'' solutions result from the sparseness of the ROSAT and
Chandra observations. (b) Same as for Figure 1a, but in the vicinity
of the $\dot\nu = 6\times 10^{-16}$ Hz s$^{-1}$ solution of Hakala et
al. (2003). This solution has a better fit with $\chi^2_{min} = 33.1$.
\label{f1}}

\figcaption[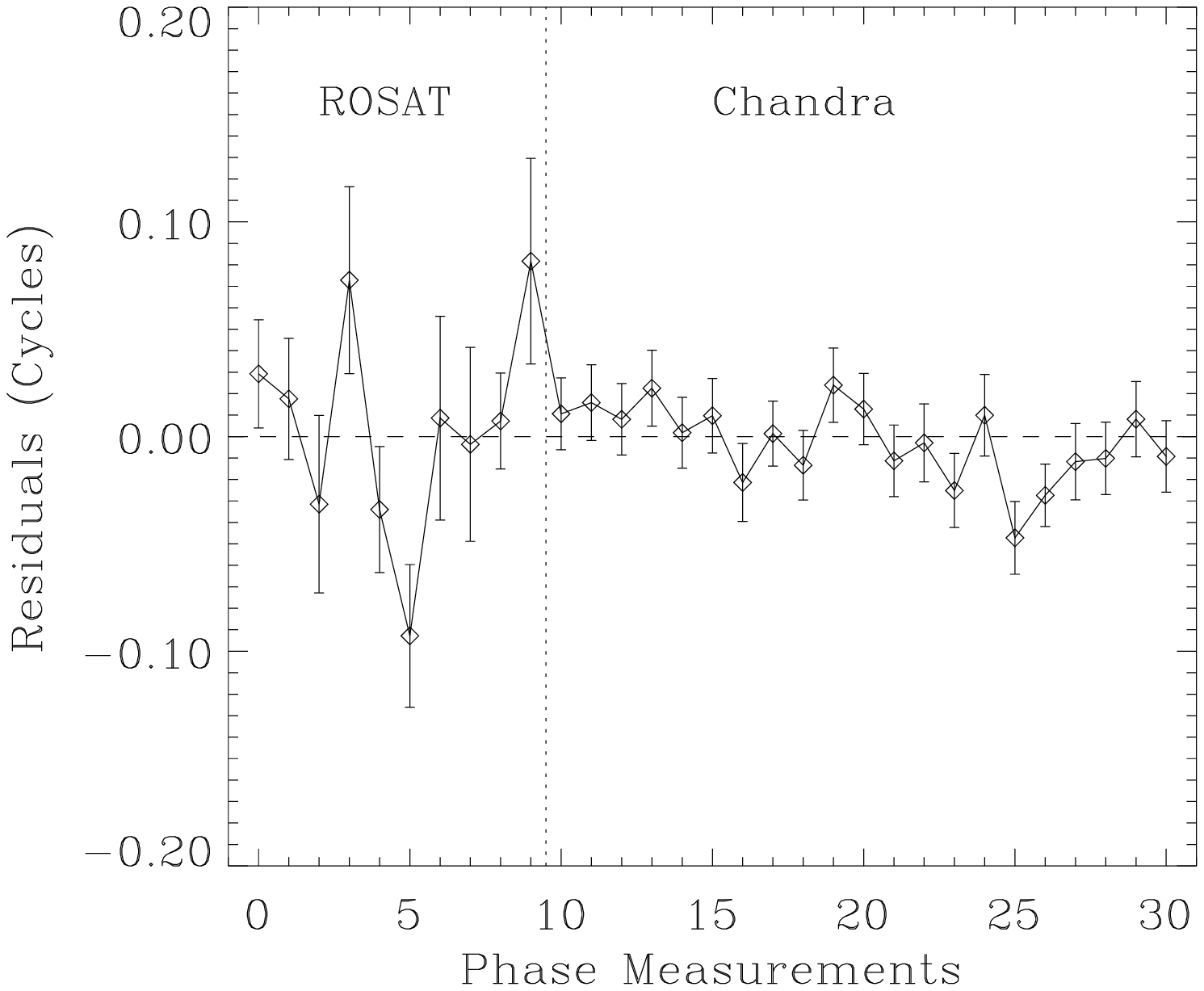, 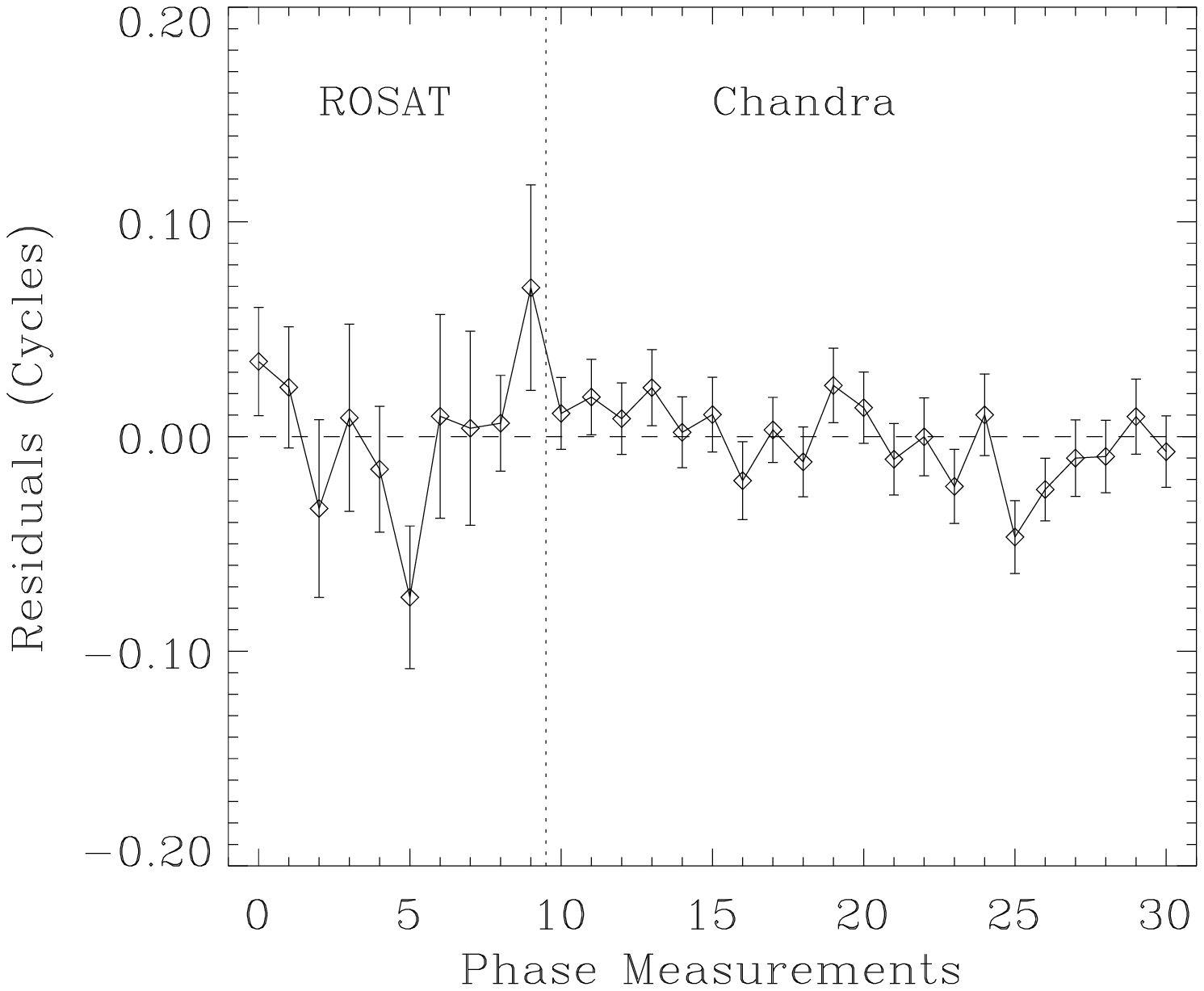]{Phase residuals from our ROSAT -- Chandra
timing study for J0806.  (a) The phase residuals from the best fit
consistent with the $\dot\nu \approx 3\times 10^{-16}$ Hz s$^{-1}$
solution of Hakala et al. (2003) are shown.  The individual phase
measurements are on the x-axis, and the vertical dotted line denotes
the ROSAT and Chandra epochs.  The long time gaps are essentially
removed in this representation. (b) Same as Figure 2a, but for the
$\dot\nu \approx 6\times 10^{-16}$ Hz s$^{-1}$ solution of Hakala et
al. (2003). This solution has the better overall $\chi^2$. Most of the
improvement comes from the better fit to the ROSAT phases.
\label{f2}}

\figcaption[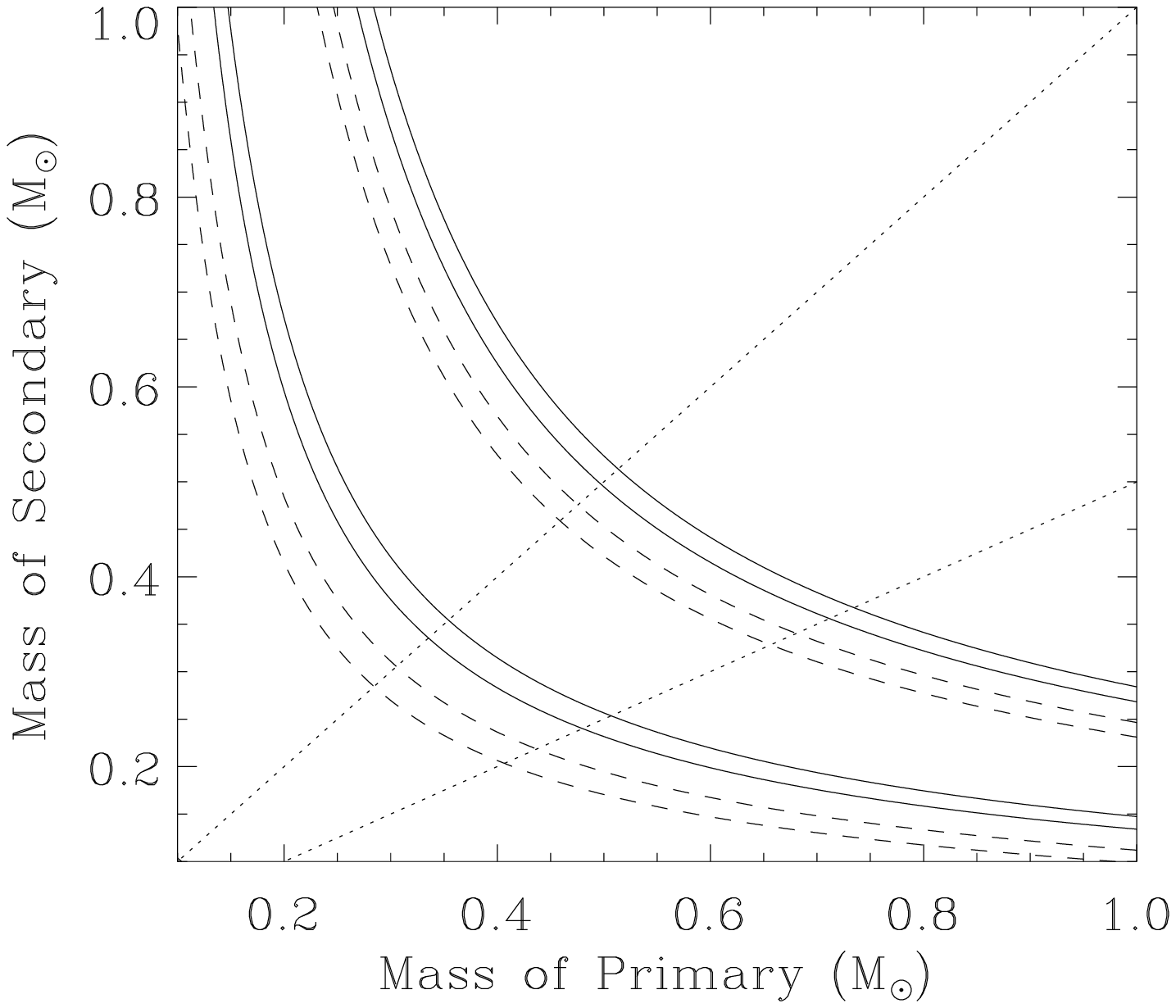]{Constraints on the component masses of J0806
assuming that the orbital decay results only from gravitational
radiation losses. The $1\sigma$ constraints for the $6 \times
10^{-16}$ Hz s$^{-1}$ (upper) and $3 \times 10^{-16}$ Hz s$^{-1}$
(lower) are shown.  In each case the solid and dashed curves are
derived assuming no spin - orbit coupling, and full synchronization,
respectively. The region between the dotted lines marks the phase
space for which $0.5 < q = m_{sec}/m_{prim} < 1.0$.
\label{f3}}

\pagebreak

\begin{figure}
\begin{center}
 \includegraphics[width=6in, height=6in]{f1a.ps}
\end{center}
Figure 1a: Results of our ROSAT -- Chandra timing study for J0806. (a)
The $1\sigma$ confidence contours in the $\nu$, $\dot\nu$ plane are
shown in the vicinity of the lower $\dot\nu \approx 3\times 10^{-16}$
Hz s$^{-1}$ solution of Hakala et al. (2003). The best $\chi^2$ for
this solution was $\chi^2_{min} = 40.8$. The Hakala et al. solution
and $1\sigma$ errors are marked by the horizontal solid and dashed
lines, respectively. The nearby, ``alias'' solutions result from the
sparseness of the ROSAT and Chandra observations. (b) Same as for
Figure 1a, but in the vicinity of the $\dot\nu = 6\times 10^{-16}$ Hz
s$^{-1}$ solution of Hakala et al. (2003). This solution has a better
fit with $\chi^2_{min} = 33.1$.
\end{figure}
\clearpage

\begin{figure}
\begin{center}
\includegraphics[width=6in, height=6in]{f1b.ps}
\end{center}
Figure 1b:
\end{figure}
\clearpage

\begin{figure}
\begin{center}
 \includegraphics[width=6in, height=6in]{f2a.ps}
\end{center}

Figure 2a: Phase residuals from our ROSAT -- Chandra timing study for
J0806.  (a) The phase residuals from the best fit consistent with the
$\dot\nu \approx 3\times 10^{-16}$ Hz s$^{-1}$ solution of Hakala et
al. (2003) are shown.  The individual phase measurements are on the
x-axis, and the vertical dotted line denotes the ROSAT and Chandra
epochs.  The long time gaps are essentially removed in this
representation. (b) Same as Figure 2a, but for the $\dot\nu \approx
6\times 10^{-16}$ Hz s$^{-1}$ solution of Hakala et al. (2003). This
solution has the better overall $\chi^2$. Most of the improvement
comes from the better fit to the ROSAT phases.
\end{figure}
\clearpage

\begin{figure}
\begin{center}
 \includegraphics[width=6in, height=6in]{f2b.ps}
\end{center}
Figure 2b: 
\end{figure}
\clearpage

\begin{figure}
\begin{center}
 \includegraphics[width=6in, height=6in]{f3.ps}
\end{center}
Figure 3: Constraints on the component masses of J0806 assuming that
the orbital decay results only from gravitational radiation
losses. The $1\sigma$ constraints for the $6 \times 10^{-16}$ Hz
s$^{-1}$ (upper) and $3 \times 10^{-16}$ Hz s$^{-1}$ (lower) solutions
are shown.  In each case the solid and dashed curves are derived
assuming no spin - orbit coupling, and full synchronization,
respectively. The region between the dotted lines marks the phase
space for which $0.5 < q = m_{sec}/m_{prim} < 1.0$.
\end{figure}

\clearpage

%%%%%%%%%%%%%%%%%%%%%%%%%%%
%%%%% End of document %%%%%
%%%%%%%%%%%%%%%%%%%%%%%%%%%

\end{document}